\documentclass[a4paper,11pt]{article}
\pdfoutput=1 

\usepackage{jcappub} 

\usepackage[T1]{fontenc} 

\title{\boldmath Cavity-mode couplings in axion dark matter searches}

\author[a]{Byeong Rok Ko}
\affiliation[a]{Department of Accelerator Science, Korea University Sejong Campus, Sejong 30019, Republic of Korea}
\emailAdd{brko@korea.ac.kr}

\abstract{
  Axion dark matter searches use a microwave cavity for the
  resonant conversion of axions into photons to enhance experimental
  sensitivity, with the cavity generally configured as a two-port
  system for both signal pickup and cavity characterization
  measurements.
  In this study, we investigated cavity-mode couplings in such a
  two-port system and examined their impact on axion dark matter
  search experiments, which typically use one strongly coupled port
  and one weakly coupled port.  
  We found that, in such a two-port cavity system, the measured
  coupling strength of one port depends on that of the other; hence,
  the coupling coefficients appearing in the relation for the unloaded
  quality factor of the cavity mode can vary substantially with the
  measured coupling strengths.
  Meanwhile, the scanning rate, the figure-of-merit for axion dark
  matter searches, cancels the systematic contribution from the
  strongly coupled port; hence, the remaining systematic uncertainty
  arises only from the weakly coupled port and may be negligible,
  depending on its coupling strength.
  Nevertheless, we recommend measuring the coupling strength of the
  weakly coupled port to eliminate this systematic uncertainty and
  thereby recover any experimental sensitivity that may have been
  lost, for example by approximately 10\% when the coupling strength
  of the weakly coupled port is 0.05. 
}

\begin{document}
\maketitle
\flushbottom

\section{Introduction}\label{sec:intro}
Cosmological measurements provide compelling evidence for the
existence of dark matter (DM) in the
Universe~\cite{CDM-EVIDENCE1,CDM-EVIDENCE2}; in particular, the standard
model of Big Bang cosmology and high-precision cosmological
measurements indicate that DM accounts for approximately 85\% of the
total matter content~\cite{PLANCK}.
However, the Standard Model of particle physics, the current best
description of modern particle physics, fails to explain both the
properties and existence of DM. In light of the present galaxy
formation, DM must be cold, massive, and stable. The
axion~\cite{AXION1,AXION2} therefore remains one of the leading cold
DM candidates because it satisfies these requirements and is hence
referred to as axion DM. Originally, the axion was proposed as a
natural solution to the strong charge-conjugation and parity problem
in the Standard Model of particle
physics~\cite{strongCP1,strongCP2,strongCP3,strongCP4,strongCP5}. Within
this framework, Peccei and Quinn introduced a new symmetry~\cite{PQ},
whose spontaneous breaking generates the axion. Under these
considerations, axion DM is a compound term encompassing particle
physics and cosmology and has therefore motivated numerous
experimental searches.

At present, the axion haloscope remains the most sensitive instrument
for detecting axion DM in the microwave region because it relies on
the resonant conversion of axions into photons through axion-photon
coupling in a microwave resonant cavity, as proposed by
Sikivie~\cite{sikivie-PRL,sikivie-PRD}.
By pushing experimental parameters to their limits, recent axion
haloscope
searches~\cite{ADMX-DFSZ1,ADMX-DFSZ2,ADMX-DFSZ3,ADMX-DFSZ4,12TB-PRL,12TB-PRX,12TB-PRL2}
have achieved sensitivity to the Dine-Fischler-Srednicki-Zhitnitskii
(DFSZ) axion~\cite{DFSZ1,DFSZ2} under the assumptions that axions
constitute 100\% of the local DM density~\cite{DMRHO1, DMRHO2} and
follow the standard halo model~\cite{SHM}. Because the DFSZ axion can
be implemented in grand unified theories~\cite{GUT}, improving
experimental sensitivity and demonstrating that cold DFSZ axion DM
constitutes 100\% of the local DM density would mark a major milestone.

Against this background, we investigated cavity-mode couplings in a
two-port system and evaluated their contributions to axion DM search
experiments, which typically involve one strongly coupled port and one
weakly coupled port. We found that, in such a two-port cavity system,
the measured coupling strength of one port depends on that of the
other and that the coupling terms in the relation for the
unloaded quality factor of the cavity mode can therefore vary
substantially with the measured coupling strengths.
By contrast, the scanning rate~\cite{scanrate, DOKIM}, the
figure-of-merit for axion DM searches, cancels the systematic
contribution from the coupling strength of the strongly coupled port;
hence the residual systematic uncertainty arises solely from the
coupling strength of the weakly coupled port and may be negligible
depending on its coupling strength. Accordingly, we recommend
measuring the coupling strength of the weakly coupled port to
eliminate this systematic uncertainty and thereby recover any
experimental sensitivity lost so far.
\section{Axion haloscopes with two-port cavity systems}\label{SEC:AXION}
Traditionally, axion haloscope searches have used a two-port cavity
system for both cavity performance measurements and signal extraction,
where the two-port cavity system and the relevant measurements cover
some of the same ground in ref.~\cite{TWO-PORT}.
Here, the two cavity-mode coupling coefficients $\beta_W$
and $\beta_S$ corresponding to the weakly and strongly coupled ports,
respectively, are defined as
\begin{eqnarray}
  \beta_i\equiv\frac{P_i}{P_0}=\frac{Q_0}{Q_i},
\label{EQ:Beta0}
\end{eqnarray}
where $P_0$ and $P_i$ denote the power losses at the cavity walls and
at port $i$, respectively, with $i$ representing either the weakly or
strongly coupled port.
The second equality follows from the quality factor definition
$Q_k\equiv\omega_0 U/P_k$, i.e., energy stored to energy lost per
cycle, where $Q_k$ and $P_k$ denote the quality factor and the power
loss of system component $k$, namely the cavity or a port. $U$ and
$\omega_0$ in the numerator of the $Q_k$ definition represent the
energy stored and resonant frequency of the system, respectively.
The total power loss in the system, $P_{\rm total}$, is then expressed
as the linear sum of the losses from the cavity walls and two ports
\begin{equation}
  P_{\rm total}=P_0+P_W+P_S \Leftrightarrow\frac{\omega_0 U}{Q_L}=\frac{\omega_0 U}{Q_0}+\frac{\omega_0 U}{Q_W}+\frac{\omega_0 U}{Q_S} \Leftrightarrow\frac{1}{Q_L}=\frac{1}{Q_0}+\frac{1}{Q_W}+\frac{1}{Q_S},
  \label{EQ:PTOTAL}
\end{equation}
where $Q_L=\omega_0 U/P_{\rm total}$ in the second relation denotes
the loaded quality factor of the cavity mode. Employing the quality
factor definition aforementioned, the three relations of
equation~(\ref{EQ:PTOTAL}) are equivalent to each other.
This form in equation~(\ref{EQ:PTOTAL}) shows no explicit dependence
between $P_W$ and $P_S$, thus implies that the measurements of $P_W$
and $P_S$ or equivalently $\beta_W$ and $\beta_S$ are independent of
one another, as if each were measured in the absence of the other,
that is, in a single-port system.
Accordingly, from equation~(\ref{EQ:Beta0}) and the third relation of
equation~(\ref{EQ:PTOTAL}), one can yield the unloaded quality factor
of the cavity mode, $Q_0$, in terms of $Q_L$ and the two coupling
coefficients, as shown in equation~(\ref{EQ:Q0}):
\begin{equation}
  Q_0 = Q_L(1 + \beta_W + \beta_S).
\label{EQ:Q0}
\end{equation}
Equation~(\ref{EQ:R}) yields the scanning rate
$R$~\cite{scanrate, DOKIM}
\begin{equation}
  R\propto\frac{\beta^2_S}{(1+\beta_S)^2},
\label{EQ:R}
\end{equation}
a key experimental parameter that depends only on the coupling
coefficient $\beta_S$ out of the two coupling coefficients.

Each coupling coefficient $\beta_i$ is obtained from the complex
reflection coefficient $r_i$ at port $i$ using equation~(\ref{EQ:Beta}):
\begin{subequations}
  \label{EQ:Beta}
  \begin{align}
  \beta_i&=\frac{1+|r_i|}{1-|r_i|}~~{\rm for}~\beta_i>1~({\rm over{\text -}coupled}),\label{EQ:Beta1}\\ 
         &=\frac{1-|r_i|}{1+|r_i|}~~{\rm for}~\beta_i<1~({\rm under{\text -}coupled}), \label{EQ:Beta2}
  \end{align}
\end{subequations}
where $r_i$ is a ratio of the complex voltages of the incident and
reflected waves, and $|r_i|$ is the amplitude of $r_i$.
Because $\beta_i$ in equations~(\ref{EQ:Q0}) and (\ref{EQ:R}) are
valid only at the resonant frequency, $|r_i|$ in
equation~(\ref{EQ:Beta}) is obtained at the resonant frequency. Note
that $|r_i|$ obtained far from the resonance is close to unity.
Equation~(\ref{EQ:Beta}) can also be rearranged to obtain the
amplitude of the reflection coefficient as
\begin{eqnarray}
  |r_i|=\frac{|\beta_i-1|}{\beta_i+1}
  \label{EQ:Gamma}
\end{eqnarray}
which depends only on $\beta_i$.
Therefore, equation~(\ref{EQ:Gamma}) holds for a single-port system;
accordingly, only the coupling coefficient $\beta_i$ that satisfies
this relation or is measured using a single-port system can be used in
the calculations of $Q_0$ and $R$ through equations~(\ref{EQ:Q0}) and
~(\ref{EQ:R}), respectively, as the $\beta_i$ terms in
equation~(\ref{EQ:Q0}) are independent.
The problem therefore stems from the discrepancy between the coupling
coefficient measured using a single-port system, which can be applied
in equations~(\ref{EQ:Q0}) and (\ref{EQ:R}), and that measured using
the two-port system employed in axion haloscope searches.
In simple terms, reflection measurements obtained using a single-port
system generally differ from those obtained using a two-port system.
If the measured weak-port coupling coefficient $\beta_W$ is
sufficiently small, it can be neglected. Within this limit, the system
effectively reduces to a single-port system, as typically assumed in
axion haloscope searches that rely on equations~(\ref{EQ:Q0}) and
(\ref{EQ:R}).
However, the coupling coefficients $\beta_i$ are not independent in a
two-port system, and those measured in axion haloscope searches can
differ substantially from those appropriate for
equations~(\ref{EQ:Q0}) and (\ref{EQ:R}), depending on the measured
strong-port coupling coefficient.

\section{Two-port network analysis}\label{SEC:2-port-anal}
\begin{figure}[h]
  \centering
  \includegraphics[width=0.95\columnwidth]{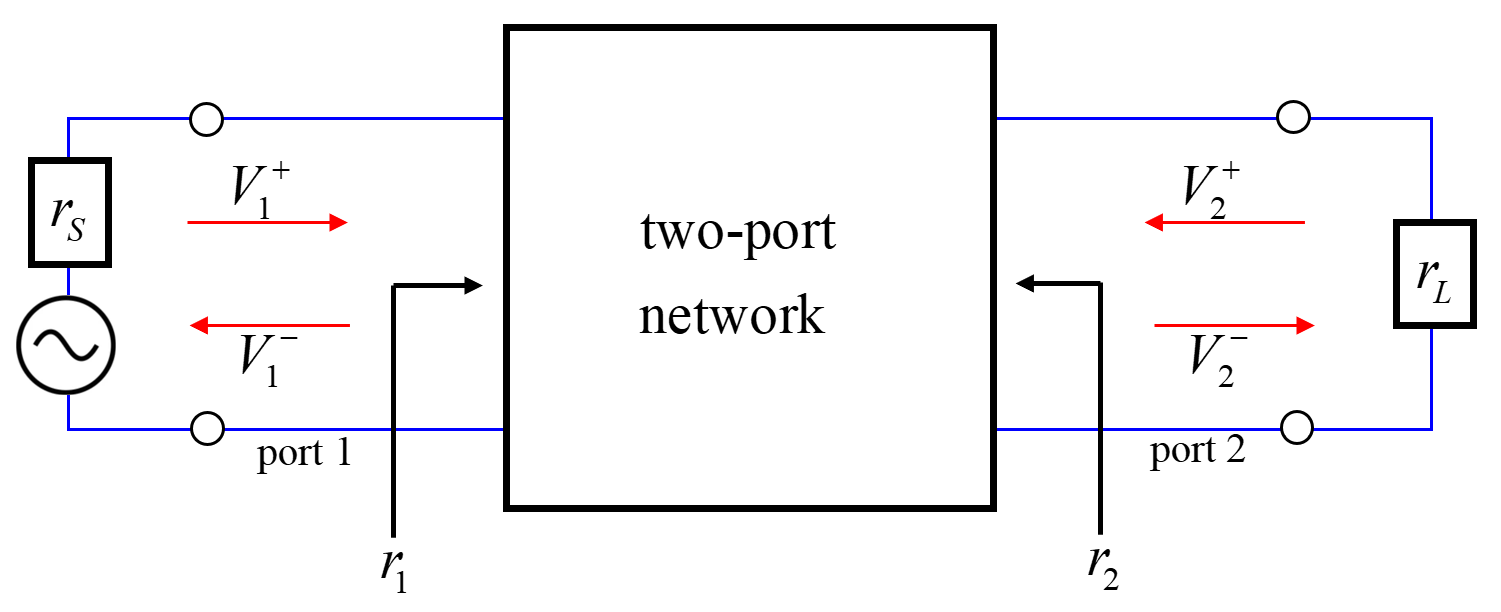}
  \caption{Diagram of a two-port network system with source reflection
    $r_S$ and load reflection $r_L$. Here, $V^+_i$ and $V^-_i$ denote
    the complex voltages of the incident and emitted waves, and
    $r_i$ the reflection coefficients, respectively, at port $i$.   
  }    
  \label{FIG:2-port}
\end{figure}

Figure~\ref{FIG:2-port} presents a two-port network diagram of the
two-port cavity system used in axion haloscope search experiments,
where the two-port network denotes the two-port cavity, source and
load correspond to the network analyzer, and blue rectangles denotes
the transmission lines.
During signal pickup, microwave switches usually disconnect the
network analyzer from the two-port network and connect port 2 to the
receiver chain, configuring a single-port measurement on a two-port
system.
Accordingly, during signal pickup, the coupling coefficient is not
determined solely by port 2, as discussed in section~\ref{SEC:AXION}
and demonstrated explicitly in this section.

The following two-port network analysis captures the interdependence
between the two ports in terms of the complex scattering parameters
$S_{ij}$~\cite{TWO-PORT,SPARM1,SPARM2}:
\begin{subequations}
  \label{EQ:SPARM}
  \begin{align}
    V^-_1&=S_{11}V^+_1+S_{12}V^+_2,\label{EQ:SPARM1}\\
    V^-_2&=S_{21}V^+_1+S_{22}V^+_2,\label{EQ:SPARM2}
  \end{align}
\end{subequations}
where $V^+_i$ and $V^-_i$ denote the complex voltages of the incident
and reflected waves, respectively, at port $i$.
The reflection coefficients at ports 1 and 2 are
$r_1=V^{-}_{1}/V^{+}_{1}$ and $r_2=V^{-}_{2}/V^{+}_{2}$, respectively.
Because $r_1$ denotes the reflection viewed from the source side,
as depicted in figure~\ref{FIG:2-port}, $V^+_2=r_L V^-_2$ with load
reflection $r_L$.
Substituting $r_L V^-_2$ for $V^+_2$ in equations~(\ref{EQ:SPARM1})
and (\ref{EQ:SPARM2}) then yields
\begin{subequations}
  \label{EQ:SPARM_NEW}
  \begin{align}
    V^-_1&=S_{11}V^+_1+S_{12}r_L V^-_2,\label{EQ:SPARM1_NEW}\\
    V^-_2&=S_{21}V^+_1+S_{22}r_L V^-_2\Leftrightarrow V^-_2=\frac{S_{21}V^+_1}{1-S_{22}r_L}.\label{EQ:SPARM2_NEW}
  \end{align}
\end{subequations}
Substituting $V^-_2$ in the second relation of
equation~(\ref{EQ:SPARM2_NEW}) to equation~(\ref{EQ:SPARM1_NEW}) and
then dividing equation~(\ref{EQ:SPARM1_NEW}) by $V^+_1$, one can yield
$r_1$ in equation~(\ref{EQ:Gamma12a}).
Meanwhile, because $r_2$ represents the reflection viewed from the
load side, as shown in figure~\ref{FIG:2-port}, $V^+_1=r_S V^-_1$ with
source reflection $r_S$.
Similar substitutions, rearrangement, and dividing done for reaching
$r_1$ in equation~(\ref{EQ:Gamma12a}), one can yield the expression for
$r_2$ in equation~(\ref{EQ:Gamma12b}).
\begin{subequations}
  \label{EQ:Gamma12}
  \begin{align}
    r_1&=\frac{V^-_1}{V^+_1}=S_{11}+\frac{S_{12}S_{21}r_L}{1-S_{22}r_L},\label{EQ:Gamma12a}\\
    r_2&=\frac{V^-_2}{V^+_2}=S_{22}+\frac{S_{21}S_{12}r_S}{1-S_{11}r_S}.\label{EQ:Gamma12b}
  \end{align}
\end{subequations}
Notably, the reflection coefficients
$r_1$ and $r_2$ in equation~(\ref{EQ:Gamma12}) correspond to
those measured in axion haloscope search experiments.
We can set $r_L=0$ and $r_S=0$ without substantial loss of
generality because most network analyzers can be calibrated well and
the transmission lines are matched to the standard 50 $\Omega$ impedance.
We simplify the analysis of $r_i$ by restricting it to the resonant
frequency, as equations~(\ref{EQ:Q0}) and (\ref{EQ:R}) are valid only
at that frequency.
\begin{subequations}
  \label{EQ:Gamma122}
  \begin{align}
    r_1&=S_{11}~~{\rm for}~r_L=0,\label{EQ:Gamma122a}\\ 
       &=\frac{1-\beta_1+\beta_2}{1+\beta_1+\beta_2}=r^*_1~~{\rm at~the~resonant~frequency}\label{EQ:Gamma122b},\\
    r_2&=S_{22}~~{\rm for}~r_S=0,\label{EQ:Gamma122c}\\ 
       &=\frac{1+\beta_1-\beta_2}{1+\beta_1+\beta_2}=r^*_2~~{\rm at~the~resonant~frequency}\label{EQ:Gamma122d},
  \end{align}
\end{subequations}
where $\beta_i$ denotes the coupling coefficient measured using a
single-port system and therefore satisfies equation~(\ref{EQ:Gamma}).
Equations~(\ref{EQ:Gamma122a}) and (\ref{EQ:Gamma122c}) follow from
setting $r_L=0$ and $r_S=0$, respectively.
While equations~(\ref{EQ:Gamma122b}) and (\ref{EQ:Gamma122d}) can be
obtained from modeling the cavity system as a damped harmonic
oscillator interacting with two external ports based on temporal
coupled-mode theory (TCMT)~\cite{TCMT1, TCMT2, TCMT3}, applying the
resonant frequency condition.
We adopted harmonic time dependence when applying TCMT, which is
appropriate for axion haloscope search experiments typically
considering a cavity mode with a quality factor of
$\mathcal{O}(10^4)$.
To distinguish single-port results from two-port results, we denote
the latter with an asterisk, as shown in the terms after the second
equalities in equations~(\ref{EQ:Gamma122b}) and
(\ref{EQ:Gamma122d}). Accordingly, $r^{*}_i$ and $r_i$ hereafter
denote the reflection coefficients of the two-port and single-port
systems, respectively.
As in axion haloscope searches, we further
assumed that port 1 is under-coupled and port 2 is over-coupled.
The amplitudes of reflection coefficients are, then,
\begin{subequations}
  \label{EQ:Gamma123}
  \begin{align}
    |r^*_1|&=|S_{11}|=\left\lvert\frac{1-\beta_1+\beta_2}{1+\beta_1+\beta_2}\right\rvert=\frac{1-\beta_1+\beta_2}{1+\beta_1+\beta_2}\label{EQ:Gamma123a},\\
    |r^*_2|&=|S_{22}|=\left\lvert\frac{1+\beta_1-\beta_2}{1+\beta_1+\beta_2}\right\rvert=\frac{-1-\beta_1+\beta_2}{1+\beta_1+\beta_2}\label{EQ:Gamma123b}.
  \end{align}
\end{subequations}
For the opposite case, the signs in the numerators of the last terms
in equation~(\ref{EQ:Gamma123}) must be reversed.
Consequently, $|r^*_i|$ no longer generally obeys
equation~(\ref{EQ:Gamma}) in a two-port network system or in axion
haloscope search experiments, but it follows equation~(\ref{EQ:Gamma})
when one of the $\beta_i$ values approaches zero, as expected.
At the resonant frequency, the transmission scattering parameters
satisfy $S_{12}=S_{21}=\frac{-2\sqrt{\beta_1\beta_2}}{1+\beta_1+\beta_2}$
~\cite{TCMT1, TCMT2, TCMT3}; hence, in a two-port network system,
adjusting $r_L$ or $r_S$ in equation~(\ref{EQ:Gamma12})
cannot render $r_i$ exclusively dependent on its port.

Using the two-port results in equation~(\ref{EQ:Gamma123}), we can
generalize equation~(\ref{EQ:Beta}):
\begin{subequations}
  \label{EQ:Beta2Port}
  \begin{align}
    \beta^{*}_1&=\frac{1-|r^{*}_1|}{1+|r^{*}_1|}=\frac{\beta_1}{1+\beta_2}~~{\rm for}~\beta^{*}_1<1~({\rm under{\text -}coupled}),\\
    \beta^{*}_2&=\frac{1+|r^{*}_2|}{1-|r^{*}_2|}=\frac{\beta_2}{1+\beta_1}~~{\rm for}~\beta^{*}_2>1~({\rm over{\text -}coupled}).
  \end{align}
\end{subequations}
As indicated in equation~(\ref{EQ:Beta2Port}), the coupling
coefficients measured in a two-port system, $\beta^*_i$, depend not
only on their own port but also on the other port.

We verified the relations in equation~(\ref{EQ:Beta2Port}) using the
finite-element simulation package, CST Studio
Suite~\cite{CST}.
Notably, the eigenmode solver in CST Studio Suite provides only the
$Q_0$ value of the cavity mode without external ports.
When external ports are added to the cavity, the solver calculates the
ports' quality factors $Q_i$'s as well as the $Q_0$ and $Q_L$ of the
cavity mode. The coupling coefficients then can be calculated as
$Q_0/Q_i$ according to equation~(\ref{EQ:Beta0}).
Because the coupling coefficients $Q_0/Q_i$, $Q_0$, and $Q_L$ obtained
from the eigenmode solver turn out to fit equation~(\ref{EQ:Q0})
well; thus, the coupling coefficient obtained from the solver
corresponds to $\beta_i$ in equation~(\ref{EQ:Beta2Port}).

By contrast, the frequency-domain solver in the finite-element
simulation provides the scattering parameters $S_{11}$, $S_{22}$,
$S_{12}$, and $S_{21}$, and the coupling coefficients can be estimated
using $S_{ii}$ and the $Q_L$ of the cavity mode derived from
$S_{ij}$. Notably, the $Q_L$ estimated from $S_{ij}$ is consistent
with that derived from the eigenmode solver, as expected, because the
frequency-domain solver simulates a network analyzer in a real
experiment and the $Q_L$ measurement correctly reflects the actual
load regardless of its impedance matching, thus is always likely
correct.
However, the $Q_0$ calculated from equation~(\ref{EQ:Q0}) by using
the coupling coefficients estimated from $S_{11}$ and $S_{22}$ differs
from that provided by the eigenmode solver, as the frequency-domain
solver mimics a network analyzer in a real experiment employing a
two-port cavity system. Hence, the coupling coefficients estimated
from $S_{11}$ and $S_{22}$ correspond to $\beta^*_1$ and $\beta^*_2$
in equation~(\ref{EQ:Beta2Port}), respectively.

Thus, CST Studio Suite also uses the relations between the coupling
coefficients $\beta^*_i$ and $\beta_i$ provided in
equation~(\ref{EQ:Beta2Port}). This constitutes another validation of
equation~(\ref{EQ:Beta2Port}).

\section{Effects on axion DM searches}\label{SEC:EFFECTS}
In the discussion in section~\ref{SEC:2-port-anal}, ports 1 and 2 were
considered under-coupled and over-coupled, respectively; thus, they
can be regarded as the weakly and strongly coupled ports in axion
haloscope search experiments.
Then, $\beta^{*}_1=\beta^{*}_W=\beta_W/(1+\beta_S)$ and
$\beta^{*}_2=\beta^{*}_S=\beta_S/(1+\beta_W)$ from equation~(\ref{EQ:Beta2Port}),
and $\beta_W$ and $\beta_S$ can be derived as
\begin{subequations}
  \label{EQ:Beta2Q0}
  \begin{align}
    \beta_1&=\beta_W=\frac{\beta^{*}_W(1+\beta^{*}_S)}{1-\beta^{*}_W\beta^{*}_S},\\
    \beta_2&=\beta_S=\frac{\beta^{*}_S(1+\beta^{*}_W)}{1-\beta^{*}_W\beta^{*}_S}.
  \end{align}
\end{subequations}
These expressions can be used in the calculations of
equation~(\ref{EQ:Q0}), with $\beta^{*}_W\beta^{*}_S<1$ which is
the typical coupling combination of axion dark matter search
experiments.
Reference~\cite{MULTIPORT} reports a closely related result
using different methods and also presents more general coupling
outcomes.
\begin{figure}[h]
  \centering
  \includegraphics[width=0.95\columnwidth]{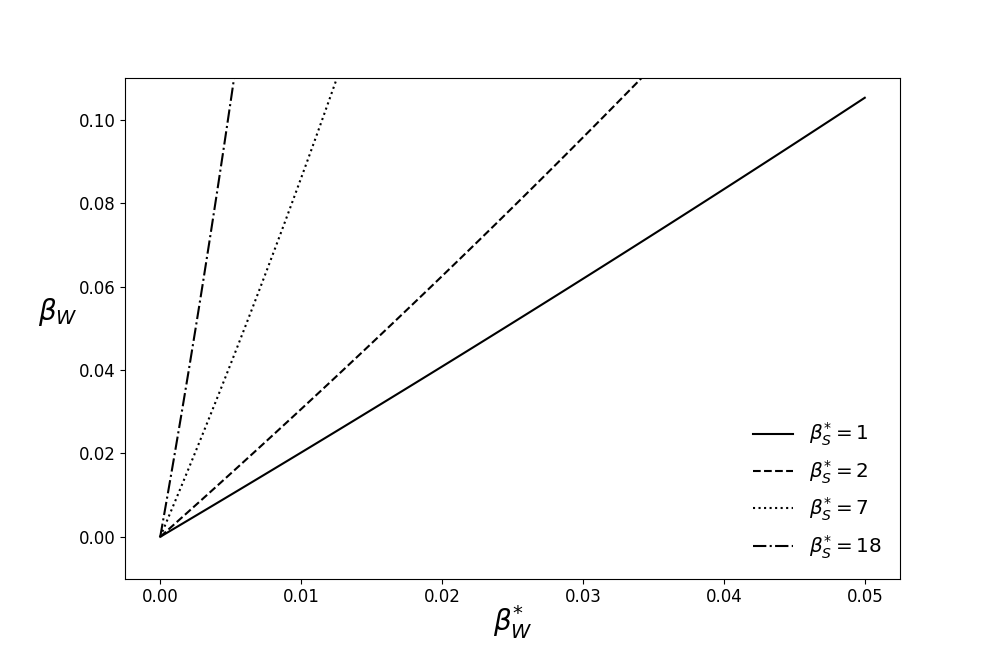}
  \caption{$\beta_W$ as a function of $\beta^*_W$ for different $\beta^*_S$ values.}  
  \label{FIG:BetaW}
\end{figure}

Most axion haloscope searches employ $\beta^{*}_S\ge1$, which implies
that $\beta_W$ is at least $2\beta^{*}_W$ and may depart considerably
from it depending on the measured value of
$\beta^{*}_S$. Figure~\ref{FIG:BetaW} presents $\beta_W$ as a function
of $\beta^{*}_W$ for four $\beta^{*}_S$ values.
Notably, $\beta^{*}_S=$1 and 2 are typical choices in axion
haloscope searches to maximize the signal-to-noise ratio and scanning
rate, respectively.
Substantially larger $\beta^{*}_S$ values of approximately seven and
18 have been considered in refs.~\cite{HAYSTAC_NATURE}
and~\cite{DOKIM}, respectively.
As illustrated in figure~\ref{FIG:BetaW}, the $\beta_W$ value used to
estimate $Q_0$ must be corrected based on the measured $\beta^{*}_W$
unless $\beta^{*}_W$ is very small. This correction becomes essential
for relatively large $\beta^{*}_S$ because $\beta_W$ increases much
more rapidly than $\beta^{*}_W$.

\begin{figure}[h]
  \centering
  \includegraphics[width=0.95\columnwidth]{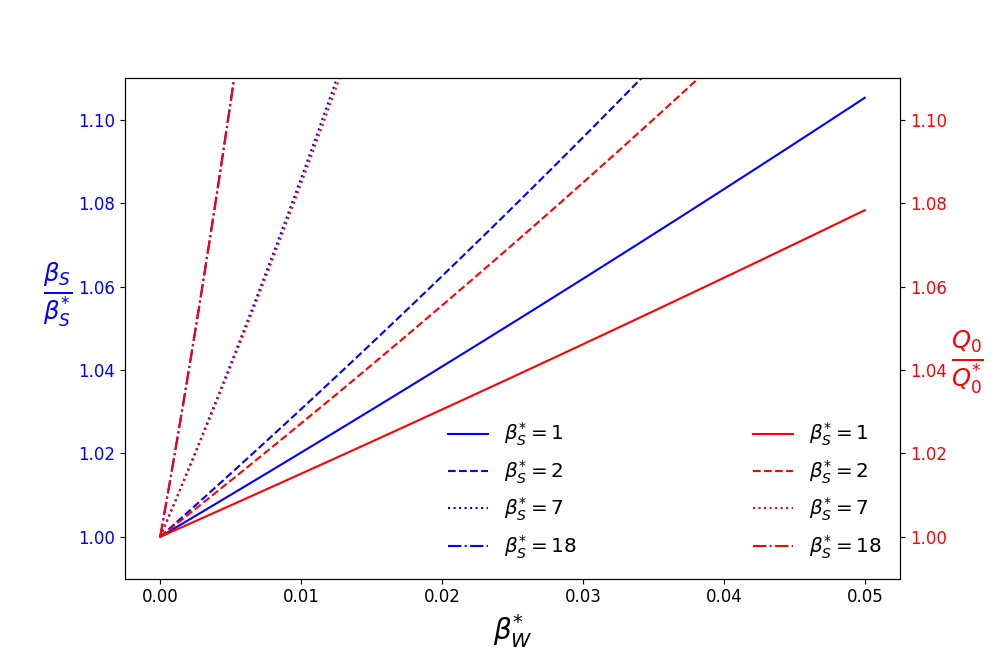}
  \caption{Blue lines with different line styles depict $\beta_S$
    normalized by $\beta^*_S$ as a function of $\beta^*_W$ for
    different $\beta^*_S$. Red lines show the corresponding curves for
    $Q_0$ normalized by $Q^*_0$.}  
  \label{FIG:RRatios}
\end{figure}

Because $\beta_S$ depends on the measured value of $\beta^{*}_W$, the
difference between $\beta_S$ and $\beta^{*}_S$ is less pronounced than
that in the $\beta_W$ case. To display this difference clearly, we
plotted $\beta_S$ as a function of $\beta^{*}_W$ and normalized it by
$\beta^{*}_S$, as shown by the blue lines with different line styles
in figure~\ref{FIG:RRatios}. 
As depicted in this figure, the $\beta_S$ value used to estimate $Q_0$
must be corrected based on the measured $\beta^{*}_S$ value unless
$\beta^{*}_W$ is very small. Notably, for a relatively large
$\beta^*_S$, the difference between $\beta_S$ and $\beta^{*}_S$
reaches approximately 10\% even when $\beta^{*}_W$ is around 0.01,
making correction essential.

Using the expressions for $\beta_W$ and $\beta_S$ in
equation~(\ref{EQ:Beta2Q0}), $Q_0$ in equation~(\ref{EQ:Q0}) becomes
\begin{eqnarray}
  Q_0=\frac{Q_L(1+\beta^{*}_W+\beta^{*}_S+\beta^{*}_W \beta^{*}_S)}{1-\beta^{*}_W\beta^{*}_S}.
  \label{EQ:Q0New}
\end{eqnarray}
Notably, $Q_L$ is measured using a two-port system even though it does
not carry an asterisk. Equation~(\ref{EQ:Q0New}) contains the cross term
$\beta^{*}_W \beta^{*}_S$, which clearly indicates that the coupling
coefficients are not independent, unlike those in
equation~(\ref{EQ:Q0}).
We then examined how this correction affects the estimation of
$Q_0$. Notably, equation~(\ref{EQ:QRatio}) yields the $Q_0/Q^*_0$
ratio, where $Q^*_0$ denotes the unloaded quality factor of the cavity
mode without the corrections reported in this study, and
figure~\ref{FIG:RRatios} presents this ratio for different measured
$\beta^*_S$ values using red lines with different line styles.
\begin{subequations}
  \label{EQ:QRatio}
  \begin{align}
    \frac{Q_0}{Q^{*}_0}&=\frac{1+\beta_W+\beta_S}{1+\beta^{*}_W+\beta^{*}_S}=\frac{(1+\beta^{*}_W)(1+\beta^{*}_S)}{(1-\beta^{*}_W\beta^{*}_S)(1+\beta^{*}_W+\beta^{*}_S)}\\
    &\simeq\frac{1+\beta^{*}_W}{1-\beta^{*}_W\beta^{*}_S}=\frac{\beta_S}{\beta^{*}_S}~~{\rm for~a~large~}\beta^{*}_S\gg1~{\rm or~}1+\beta^{*}_W+\beta^{*}_S\simeq1+\beta^{*}_S.
  \end{align}
\end{subequations}
For $\beta^*_S\gg1$, equation~(\ref{EQ:QRatio}) indicates that the
quality-factor ratio approaches the ratio of the coupling strengths of
the strongly coupled port, and figure~\ref{FIG:RRatios} illustrates this
behavior. Therefore, as in the $\beta^*_S$ case, the quality factor
can differ by approximately 10\% for large $\beta^{*}_S$ even when
$\beta^{*}_W$ is only around 0.01; hence, the corrections reported
herein are essential for accurate $Q_0$ estimation and may be relevant
to the cases reported in refs.~\cite{SRF1, SRF2, SRF3}.

Finally, we examined the effect of the corrections on the scanning
rate $R$, the most important figure-of-merit in axion DM search
experiments.
\begin{eqnarray}
  \frac{R}{R^{*}}=\frac{\dfrac{\beta^2_S}{(1+\beta_S)^2}}{\dfrac{\beta^{*2}_S}{(1+\beta^{*}_S)^2}}=(1+\beta^{*}_W)^2\simeq 1+2\beta^{*}_W.
  \label{EQ:RRatio}
\end{eqnarray}
\begin{figure}[h]
  \centering
  \includegraphics[width=0.95\columnwidth]{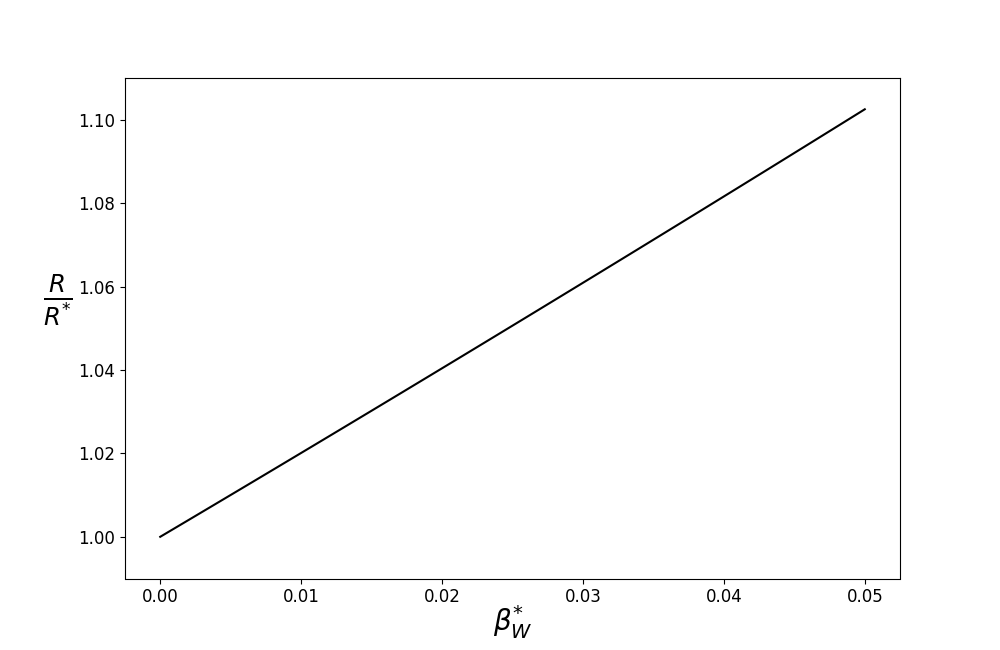}
  \caption{$R$ normalized by $R^*$ as a function of $\beta^*_W$.}
  \label{FIG:RR}
\end{figure}
Equation~(\ref{EQ:RRatio}) yields the ratio of the scanning rates with
and without the coupling coefficient corrections considered in this
study.
Fortunately, the systematic effect associated with the $\beta^*_S$
correction cancels out.
However, the $\beta^*_W$ value can still change $R$ by around 10\%
when its measured value is 0.05, as depicted in figure~\ref{FIG:RR},
even though the scanning rate depends only on the coupling coefficient
of the strongly coupled port according to equation~(\ref{EQ:R}).

\section{Summary}
In this study, we examined cavity-mode couplings in a two-port system
and evaluated their impact on axion DM search experiments.
We found that the coupling coefficients appearing in the expression for
the unloaded quality factor of the cavity mode can vary considerably
depending on the measured coupling coefficients.
Although the scanning rate is generally regarded as depending only on
the coupling coefficient of the strongly coupled port, we found that
the coupling coefficient of the weakly coupled port can also influence
the scanning rate. Unlike in the unloaded-quality-factor calculation, the
systematic contribution from the coupling coefficient of the strongly
coupled port to the scanning rate cancels out. Nevertheless, we
recommend measuring the coupling strength of the weakly coupled port
to eliminate the remaining systematic uncertainty and thereby recover
any experimental sensitivity that may otherwise be lost, for example
by approximately 10\% when the weak-port coupling coefficient is 0.05.
Finally, we mention the coupling of the weak port. To avoid any
significant loss out the weak port, it is better to have $Q_W$ of
50$\sim$100 times larger than $Q_0$ according to the third relation of
equation~(\ref{EQ:PTOTAL}), which corresponds to $\beta_W$ of
0.02$\sim$0.01 from equation~(\ref{EQ:Beta0}). Therefore, $\beta^*_W$ of
less than 0.01 is desirable in a two-port cavity system, which also
reduces the systematic uncertainty in the scanning rate stems from the
weak port coupling as shown in equation~(\ref{EQ:RRatio}).

\acknowledgments
This work was supported by a Korea University Grant, the National
Research Foundation of Korea (NRF) grants funded by the Korea
government (MSIT) (RS-2025-00556247 and RS-2022-00143178), and the
Korea Basic Science Institute (National research Facilities and
Equipment Center) grant funded by the Korea government (MSIT)
(NFEC-2019R1A6C1010027).


\begin{thebibliography}{99}

\bibitem{CDM-EVIDENCE1}
  V. C. Rubin and W. K. Ford Jr., \emph{Rotation of the Andromeda nebula from a spectroscopic survey of emission regions}, \emph{ApJ} \textbf{159} (1970) 379.
\bibitem{CDM-EVIDENCE2}
  Douglas Clowe \emph{et al.}, \emph{A direct empirical proof of the existence of dark matter}, \emph{ApJ} \textbf{648} (2006) L109.

\bibitem{PLANCK}
  P. A. R. Ade \emph{et al.} (Planck Collaboration), \emph{Planck 2015 results. XIII. Cosmological parameters}, \emph{Astron. Astrophys.} \textbf{594} (2016) A13. 
  
\bibitem{AXION1}
  S. Weinberg, \emph{A new light boson?}, \emph{Phys. Rev. Lett.} \textbf{40} (1978) 223.
\bibitem{AXION2}
  F. Wilczek, \emph{Problem of strong P and T invariance in the presence of instantons}, \emph{Phys. Rev. Lett.} \textbf{40} (1978) 279.

\bibitem{strongCP1}
  G. 't Hooft, \emph{Symmetry breaking through Bell–Jackiw anomalies}, \emph{Phys. Rev. Lett.} {\bf 37} (1976) 8.
\bibitem{strongCP2}
  G. 't Hooft, \emph{Computation of the quantum effects due to a four-dimensional pseudoparticle}, \emph{Phys. Rev. D} {\bf 14} (1976) 3432 [Erratum ibid. {\bf 18} (1978) 2199].
\bibitem{strongCP3}
  J. H. Smith, E. M. Purcell, and N. F. Ramsey, \emph{Experimental limit to the electric dipole moment of the neutron}, \emph{Phys. Rev.} \textbf{108} (1957) 120.
\bibitem{strongCP4}
  W. B. Dress, P. D. Miller, J. M. Pendlebury, P. Perrin, and
  N. F. Ramsey, \emph{Search for an electric dipole moment of the neutron}, \emph{Phys. Rev. D} {\bf 15} (1977) 9.
\bibitem{strongCP5}
  I. S. Altarev \emph{et al.}, \emph{A search for the electric dipole moment of the neutron using ultracold neutrons}, \emph{Nucl. Phys.} \textbf{A341} (1980) 269. 
  
\bibitem{PQ}
  R. D. Peccei and H. R. Quinn, \emph{CP conservation in the presence of pseudoparticles}, \emph{Phys. Rev. Lett.} \textbf{38} (1977) 1440.

\bibitem{sikivie-PRL}
  P. Sikivie, \emph{Experimental tests of the “invisible” axion}, \emph{Phys. Rev. Lett.} \textbf{51} (1983) 1415 [Erratum ibid. {\bf 52} (1984) 695].

\bibitem{sikivie-PRD}
  P. Sikivie, \emph{Detection rates for “invisible”-axion searches}, \emph{Phys. Rev. D} \textbf{32} (1985) 2988 [Erratum ibid. {\bf 36} (1987) 974].

\bibitem{ADMX-DFSZ1}
  N. Du \emph{et al.} (ADMX Collaboration), \emph{A search for invisible axion dark matter with the Axion Dark Matter Experiment}, \emph{Phys. Rev. Lett.} \textbf{120} (2018) 151301.
\bibitem{ADMX-DFSZ2}
  T. Braine \emph{et al.} (ADMX Collaboration), \emph{Extended search for the invisible axion with the Axion Dark Matter Experiment}, \emph{Phys. Rev. Lett.} \textbf{124} (2020) 101303.
\bibitem{ADMX-DFSZ3}
  C. Bartram \emph{et al.} (ADMX Collaboration), \emph{Search for invisible axion dark matter in the 3.3–4.2 $\mu$eV mass range}, \emph{Phys. Rev. Lett.} \textbf{127} (2021) 261803.
\bibitem{ADMX-DFSZ4}
  C. Goodman \emph{et al.} (ADMX Collaboration), \emph{ADMX axion dark matter bounds around 3.3 $\mu$eV with Dine-Fischler-Srednicki-Zhitnitsky discovery ability}, \emph{Phys. Rev. Lett.} \textbf{134} (2025) 111002.
  
\bibitem{12TB-PRL}
  Andrew K. Yi \emph{et al.}, \emph{Axion dark matter search around 4.55 $\mu$eV with Dine–Fischler–Srednicki–Zhitnitskii sensitivity}, \emph{Phys. Rev. Lett.} \textbf{130} (2023) 071002.

\bibitem{12TB-PRX}
  Saebyeok Ahn \emph{et al.}, \emph{Extensive search for axion dark matter over 1 GHz with CAPP’s main axion experiment}, \emph{Phys. Rev. X} \textbf{14} (2024) 031023.

\bibitem{12TB-PRL2}
  Saebyeok Ahn \emph{et al.}, \emph{Extended Haloscope Search and Exclusion of a Candidate Signal near 1.036~GHz}, \emph{Phys. Rev. Lett.} \textbf{137} (2026) 021803.
  
\bibitem{DFSZ1}
  A. R. Zhitnitskii, \emph{On possible suppression of the axion hadron interactions}, \emph{Yad. Fiz.} {\bf 31} (1980) 497 [\emph{Sov. J. Nucl. Phys.} \textbf{31} (1980) 260].
\bibitem{DFSZ2}
  M. Dine, W. Fischler, and M. Srednicki, \emph{A simple solution to the strong CP problem with a harmless axion}, \emph{Phys. Lett.} \textbf{104B} (1981) 199.
  
\bibitem{DMRHO1}
   P. Salucci, F. Nesti, G. Gentile, and C. Frigerio Martins, \emph{The dark matter density at the Sun’s location}, \emph{A\&A} 523 (2010) A83.
   
\bibitem{DMRHO2}
  Fabrizio Nesti and Paolo Salucci, \emph{The dark matter halo of the Milky Way}, \emph{JCAP} 07 (2013) 016.

\bibitem{SHM}
  M. S. Turner, \emph{Periodic signatures for the detection of cosmic axions}, \emph{Phys. Rev. D} \textbf{42} (1990) 3572.

\bibitem{GUT}
  Howard Georgi and S. L. Glashow, \emph{Unity of all elementary-particle forces}, \emph{Phys. Rev. Lett.} \textbf{32} (1974) 438.

\bibitem{scanrate}
  L. Krauss, J. Moody, F. Wilczek, and D. E. Morris, \emph{Calculations for cosmic axion detection}, \emph{Phys. Rev. Lett.} \textbf{55} (1985) 1797.

\bibitem{DOKIM}
  Dongok Kim \emph{et al.}, \emph{Revisiting the detection rate for axion haloscopes}, \emph{JCAP} {\bf 03} (2020) 066.

\bibitem{TWO-PORT}
  Dae-Hyun Han, Young-Soo Kim, and M. Kwon, \emph{Two port caviyt Q measurement using scattering parameters}, \emph{Rev. Sci. Instrum.} {\bf 67} (1996) 2179.  

\bibitem{SPARM1}
  D. M. Pozar (2011). Microwave engineering, 4th ed., Wiley, Hoboken U.S.A. (2011).

\bibitem{SPARM2}
  F. Gustrau (2025). In RF and microwave engineering:Fundamentals of wireless communications, 4th ed., Wiley, Hoboken U.S.A. (2025).
  
\bibitem{TCMT1}
  S. Fan, W. Suh, and J. D. Joannopoulos, \emph{Temporal coupled-mode theory for the Fano resonance in optical resonators}, \emph{J. Opt. Soc. Am. A} {\bf 20}(3) (2003) 569-572.

\bibitem{TCMT2}
  S. Fan, Photonic crystal theory: Temporal coupled-mode formalism in Photonic Crystals: Physics, Fabrication and Applications, edited by K. Inoue and K. Ohtaka, Springer, Berlin;New York (2004), pp. 431-456.  

\bibitem{TCMT3}
  Z. Zhao, C. Guo, and S. Fan, \emph{Connection of temporal coupled-mode-theory formalisms for a resonant optical system and its time-reversal conjugate}, \emph{Phys. Rev. A} {\bf 99}(3) (2019) 033839.

\bibitem{CST}
  CST Studio Suite, Dassault Syst$\grave{\rm e}$mes. \url{https://www.3ds.com/products/simulia/cst-studio-suite}.

\bibitem{MULTIPORT}
  D. Frolov, \emph{Intrinsic quality factor extraction of multi-port cavity with arbitrary coupling}, \emph{Rev. Sci. Instrum.} \textbf{92} (2021) 014704.

\bibitem{HAYSTAC_NATURE}
  K. M. Backes \emph{et al.}, \emph{A quantum enhanced search for dark matter axions}, \emph{Nature} {\bf 590} (2021) 238–242.


\bibitem{SRF1}
  D. Alesini \emph{et al.}, \emph{Galactic axions search with a superconducting resonant cavity}, \emph{Phys. Rev. D} \textbf{99} (2019) 101101(R).

\bibitem{SRF2}
  Danho Ahn \emph{et al.}, \emph{Biaxially textured $\text{YBa}_{2}\text{Cu}_{3}\text{O}_{7-x}$ microwave cavity in a high magnetic field for a dark-matter axion search}, \emph{Phys. Rev. A} \textbf{17} (2022) L061005.

\bibitem{SRF3}
  S. Ahyoune \emph{et al.}, \emph{RADES axion search results with a high-temperature superconducting cavity in an 11.7 T magnet}, \emph{J. High Energ. Phys.} \textbf{04} (2025) 113.
    
\end{thebibliography}
\end{document}